# Formation of Droplets of the Order Parameter and Superconductivity in Inhomogeneous Fermi–Bose Mixtures (Brief Review)

**M. Yu. Kagan** [a,b,*], **S. V. Aksenov** [c], **A. V. Turlapov** [d,e,f], **R. Sh. Ikhsanov** [a], **K. I. Kugel** [g,a], **E. A. Mazur** [h,i], **E. A. Kuznetsov** [j,k,l], **V. M. Silkin** [m,n,o], and **E. A. Burovski** [a]

[a] *National Research University Higher School of Economics, Moscow, 101000 Russia*

[b] *Kapitza Institute for Physical Problems, Russian Academy of Sciences, Moscow, 119334 Russia*

[c] *Kirensky Institute of Physics, Federal Research Center KSC, Siberian Branch, Russian Academy of Sciences, Krasnoyarsk, 660036 Russia*

[d] *Institute of Applied Physics, Russian Academy of Sciences, Nizhny Novgorod, 603950 Russia*

[e] *Moscow Institute of Physics and Technology (National Research University), Dolgoprudnyi, Moscow region, 141700 Russia*

[f] *Russian Quantum Center, Skolkovo, Moscow, 143025 Russia*

[g] *Institute for Theoretical and Applied Electrodynamics, Russian Academy of Sciences, Moscow, 125412 Russia*

[h] *National Research Nuclear University MEPhI (Moscow Engineering Physics Institute), Moscow, 115409 Russia*

[i] *National Research Center Kurchatov Institute, Moscow, 123182 Russia*

[j] *Lebedev Physical Institute, Russian Academy of Sciences, Moscow, 119991 Russia*

[k] *Landau Institute of Theoretical Physics, Russian Academy of Sciences, Chernogolovka, Moscow region, 142432 Russia*

[l] *Skolkovo Institute of Science and Technology, Skolkovo, Moscow, 121205 Russia*

[m] *Donostia International Physics Center (DIPC), San Sebastian/Donostia, Basque Country, 20018 Spain*

[n] *Departamento de Polimeros y Materiales Avanzados: Fisica, Quimica y Tecnologia, Facultad de Ciencias Quimicas, Universidad del Pais Vasco UPV/EHU, Apartado 1072, San Sebastiain/Donostia, Basque Country, 20080 Spain*

[o] *IKERBASQUE, Basque Foundation for Science, Bilbao, Basque Country, 48013 Spain*

[*] *e-mail: kagan@kapitza.ras.ru*

**Abstract**

The studies of a number of systems treated in terms of an inhomogeneous (spatially separated) Fermi–Bose mixture with superconducting clusters or droplets of the order parameter in a host medium with unpaired normal states are reviewed. A spatially separated Fermi–Bose mixture is relevant to superconducting BaKBiO$_3$ bismuth oxides. Droplets of the order parameter can occur in thin films of a dirty metal, described in the framework of the strongly attractive two-dimensional Hubbard model at a low electron density with a clearly pronounced diagonal disorder. The Bose–Einstein condensate droplets are formed in mixtures and dipole gases with an imbalance in the densities of the Fermi and Bose components. The Bose–Einstein condensate clusters also arise at the center or at the periphery of a magnetic trap involving spin-polarized Fermi gases. Exciton and plasmon collapsing droplets can emerge in the presence of the exciton–exciton or plasmon–plasmon interaction. The plasmon contribution to the charge screening in MgB$_2$ leads to the formation of spatially modulated inhomogeneous structures. In metallic hydrogen and metal hydrides, droplets can be formed in shock-wave experiments at the boundary of the first-order phase transition between the metallic and molecular phases. In a spatially separated Fermi–Bose mixture arising in an Aharonov–Bohm interference ring with a superconducting bridge in a topologically nontrivial state, additional Fano resonances may appear and collapse due to the presence of edge Majorana modes in the system.

## 1. INTRODUCTION. SPATIALLY SEPARATED FERMI–BOSE MIXTURE AND SUPERCONDUCTIVITY IN BaKBiO$_3$ BISMUTH OXIDES

The idea of a spatially separated Fermi–Bose mixture of finite size boson clusters or complexes (containing compact electron or hole pairs) in a host medium involving unpaired (Fermi liquid) states was first formulated in [1] with the aim to explain the mechanism of superconductivity and the nature of the electron transport in the normal state of BaKBiO$_3$ bismuth oxides.

Figure 1 shows (left panel) the diagram illustrating the local crystal structure in the BiO$_2$ plane of the parent compound BaBiO$_3$, which is an insulator with a charge density wave (CDW) and the checkerboard structure of the distribution of BiL$^2$O$_6$ and BiO$_6$ octahedra, as well as (right panel) the diagram illustrating the local crystal structure of Ba$_{0.5}$K$_{0.5}$BiO$_3$ with diagonal chains of BiO$_6$ complexes embedded in a large percolation cluster formed by BiL$^2$O$_6$ octahedra.

We note that the formation of the long-range order and of the macroscopic wavefunction characteristic of the superconducting state in Ba$_{1-x}$K$_x$BiO$_3$ is due to the tunneling of local electron pairs from one $BiO_6$ boson cluster to the neighboring $BiO_6$ cluster through the tunneling barrier formed by normal fermion clusters. The components of the Fermi–Bose mixture at the concentration of metal in the range of $0.37 < x < 0.5$ are separated in the real space, but not in the energy space (see [1, 2] for a more detailed description).

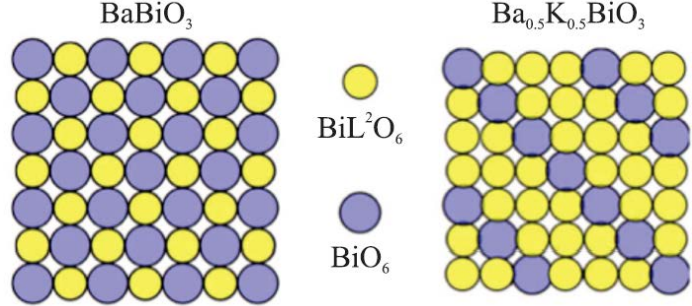

**Fig. 1.** Schematic of (left panel) the local crystal structure in the $BiO_2$ plane of the parent $BaBiO_3$ compound, which is a CDW type insulator with a checkerboard array formed by $BiL^2O_6$ and $BiO_6$ octahedra, and (right panel) the local crystal structure of $Ba_{0.5}K_{0.5}BiO_3$ with diagonal chains of $BiO_6$ complexes embedded in a large percolation cluster formed by $BiL^2O_6$ octahedra [2].

Note also that the formation of ferromagnetic metallic nanodroplets in nonmagnetic hosts (paramagnetic, antiferromagnetic, and charge-ordered) such as insulating transition-metal oxides was discussed in detail in [3].

## 2. DROPLETS OF THE ORDER PARAMETER IN A LOW-DENSITY ELECTRON SYSTEM WITH ATTRACTION IN THE PRESENCE OF A STRONG RANDOM POTENTIAL

In [4], we calculated the properties of a two-dimensional electron system with a low electron density ($n << 1$) and a strong local Hubbard attraction $|U| \gtrsim W$ ($W$ is the band width) in the presence of a strong random potential (diagonal disorder) $V$ uniformly distributed in the range from $-V$ to $+V$. Only the electron hopping to neighboring sites of a square lattice with the band width $W = 8t$ was taken into account. The calculations were carried out on a 24×24 lattice with periodic boundary conditions.

Within the Bogoliubov–de Gennes approach, the formation of inhomogeneous states was revealed in a spatially separated Fermi–Bose mixture of Cooper pairs and unpaired electrons. These states involve superconducting boson droplets of different sizes in a host with unpaired normal electronic states.

We also analyzed in detail the range of the problem parameters (such as two important ratios $|U|/W$ and $V/W$), in which the formation of a large percolation cluster begins at electron densities $n_C \approx 0.31$ and the insulator–superconductor transition occurs [4, 5].

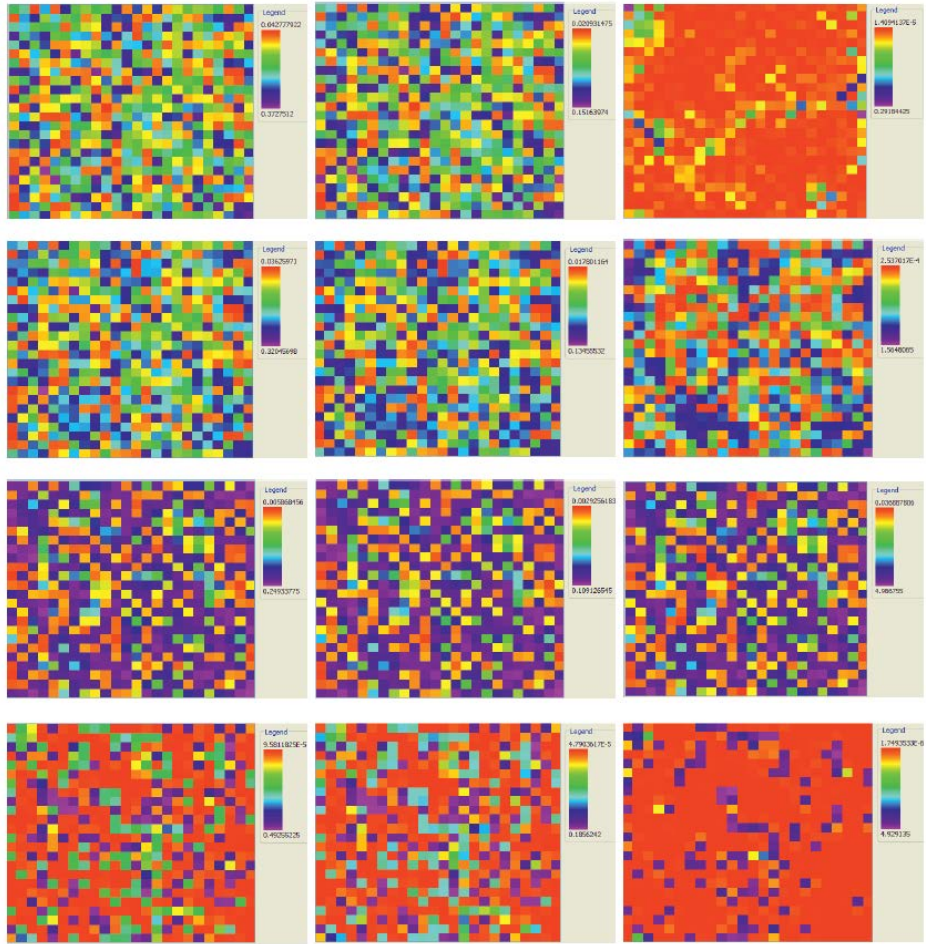

**Fig. 2.** Two-dimensional distributions of the (left panels) electron density, (middle panels) electron–hole mixing, and (right panels) order parameter at $n = 0.15$ on a $24 \times 24$ lattice with the disorder amplitude $V/t = 10.0$.

Note that the numerical simulations demonstrate the decrease in the size of droplets (from larger droplets to single electron pairs) as the electron density decreases to small values $n \leq 0.1$ at fixed values of the Hubbard attraction $U$ and random potential $V$.

The obtained results are important both for constructing the global phase diagram of our system and for understanding the nature of the phase transition between the superconducting, normal metallic, and localized (insulating) states in a quasi-two-dimensional (thin film) dirty metal. In a more practical sense, these results are also of interest for the experimentally available design of superconducting flux qubits in granular superconductors and, in particular, in granular Al films.

In [4], the calculation begins with random values of the pairing amplitude $\Delta_i$ and site-dependent renormalization of the chemical potential $\tilde{\mu}_i$ at each site, after which the numerical diagonalization of the resulting Hamiltonian was performed. As a result, the eigenvectors $u_n(\mathbf{r}_i)$ and $v_n(\mathbf{r}_i)$, as well as eigenvalues $E_n$ of the system, were determined.

In the left, middle, and right panels of Fig. 2, we show the two-dimensional distributions of the electron density, electron−hole mixing, and order parameter, respectively at $n = 0.15$ on the $24 \times 24$ lattice with the disorder amplitude $V/t = 10.0$.

We emphasize that, even for low electron densities $n \leq 0.1$ and large amplitudes of the Hubbard repulsion $|U|/t = 10$, the system is still in the Bardeen–Cooper–Schrieffer (BCS) region for the BCS−BEC crossover between extended Cooper pairs (the region of applicability of the BCS theory) and local pairs forming a Bose–Einstein condensate (BEC). Electron pairs at such parameters of the system are already quite compact and almost touch each other. Thus, the system is really very close to the limit when the pairs start to "crush" each other, i.e., to the Fermi–Bose limit of a mixture of compact pairs and single unpaired electrons. In addition, we note that, as disorder grows, the spatial distribution of the local pairing amplitude $\Delta(r)$ first takes the form of individual droplets and then, as the density decreases, of individual electron pairs.

## 3. INHOMOGENEOUS STATES IN IMBALANCED FERMI–BOSE MIXTURES AND QUANTUM GASES

The first experimental detection of the phase separation in imbalanced (spin-polarized) ultracold Fermi gases was reported in [6]. The authors of [6] detected a paired state (the Bose–Einstein condensate of compound bosons with equal densities of spin-up and spin-down components $n_\uparrow = n_\downarrow$) at the center of a three-dimensional trap, while excess up spins are mainly concentrated at the trap periphery. In quasi-one-dimensional traps, the situation is reversed: balanced paired fermions are at the periphery, and excess up spins are at the center of the trap. The first experimental results on the phase separation in quasi-two-dimensional traps were obtained in [7]. Qualitatively, the situation in the quasi-two-dimensional geometry resembles the three-dimensional one, with balanced paired fermions at the center of the trap.

In [8], studies of imbalanced ultracold gases in the dimensional crossover regime from quasi-one-dimensional to three-dimensional systems were continued. We constructed the phase diagram for a spin-polarized Fermi gas with spin-up and spin-down components in the case of an attractive interaction in the Hubbard chains and ladders with two and three legs.

The results demonstrate a significant difference between the Hubbard chains with the local on-site attraction and the two- and three-leg ladders. At the same time, the phase diagram for the two- and three-leg ladders, as well as for ladders with a larger number of legs, includes the phases of a balanced Fermi gas (ED) with local pairs at the site and the equal densities of spin-up and spin-down particles $n_\uparrow = n_\downarrow$, partially polarized Fermi gas with $n_\uparrow > n_\downarrow$ (PP), and fully polarized Fermi gas (FP) with $n_\uparrow = n$ and $n_\downarrow = 0$. The stability of the ladder systems was analyzed. The phase diagram was plotted in the convenient variables of the chemical potential $\mu$ and the magnetic field $h = n_\uparrow - n_\downarrow$.

The exact solution by the Bethe ansatz method confirms the existence of an algebraic BCS-type order in the phase with equal densities in one-dimensional systems, while the partially polarized phase is a candidate for the experimental observation of inhomogeneous superfluidity in the phase with the Fulde–Ferrell–Larkin–Ovchinnikov order parameter [9, 10], where superconducting correlations decrease algebraically in the real space and are modulated with a characteristic length determined by the difference between the Fermi momenta for the spin-up and spin-down components.

The phase diagram for the ladder systems was constructed on the basis of the Hubbard model with attraction for the isotropic case corresponding to the same values ($t_\perp = t_\parallel$) of the hopping integrals between the nearest neighbors along and across the chains (along the "legs" and rungs). We studied both the strong, $|U|/W > 1$, and weak, $|U|/W < 1$, coupling cases, where $|U|$ is the Hubbard attraction energy and $W = 4t$ is the width of the one-dimensional band.

The maximum ratio $|U|/t$ used in this study was $U/t = 7$. The calculations were carried out using the exact diagonalization technique (the density matrix renormalization group (DMRG) method) for the energy of the ground state of the system at zero temperature $T = 0$. The confinement potential in the trap was taken into account using the Thomas–Fermi approximation with replacement of the magnetic field $h$ by $h - V(x)$, and it actually corresponded to the vertical slope of the phase boundaries in the phase diagram.

Some features of the phase diagram seem interesting and important, in particular, nontrivial boundaries and nonmonotonic lines in the phase diagram between phases with partial polarization and equal numbers of particles in the case of two- and three-leg ladders. This effect becomes especially interesting near the multicritical point O on the phase diagram.

The negative slope of the boundary between these two phases, $d\mu/dh < 0$, for the Hubbard chain indicates that the partially polarized phase in the one-dimensional system corresponds to the central part of the gas cloud, while the phase with equal spin densities either is located at the periphery or is completely absent. At the same time, the slope of this line for two- and three-leg ladders becomes positive, $d\mu/dh > 0$, and the spin-polarized phase appears at the periphery of the trap, as in the experiments in [38], whereas the phase with equal spin densities is in the center of the trap.

In Fig. 3, we show the phase diagram of the Hubbard model with attraction for ladders with the number of connected chains (legs) $w$ = (a) 1, (b) 2, and (c) 3 for the on-site Hubbard attraction $U = -7t$. We show only that part of the phase diagram for which $\mu < U/2$ and $h > 0$, since its other part, with $h < 0$, is symmetric under the change $h \rightarrow -h$, and similarly, the part of the phase diagram where $\mu > U/2$ is symmetric to the one with $\mu < U/2$ owing to the particle–hole symmetry. Solid lines are the numerical calculations. The length of each leg of the ladder is $L = 40$ [8].

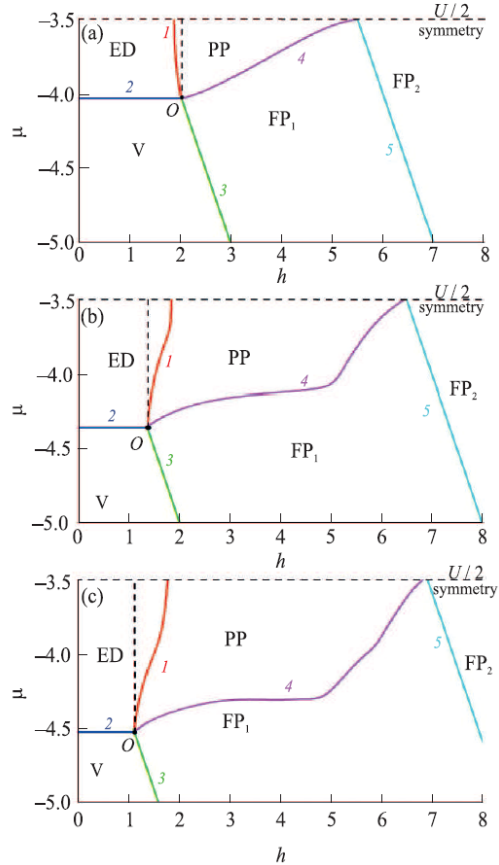

**Fig. 3.** Global phase diagram of the Hubbard model with attraction for ladders with the number of legs $w$ = (a) 1, (b) 2, and (c) 3 for the on-site Hubbard attraction $U = -7t$, where ED is the phase with equal densities of up and down spins, $n_\uparrow = n_\downarrow$ (equal density phase); PP is the partially polarized phase with spin densities $n_\uparrow > n_\downarrow$; and FP is the fully polarized phase with $n_\downarrow = 0$. We show only that part of the phase diagram for which $\mu < U/2$ and $h > 0$, since its other part, $h < 0$, is symmetric under the change $h \rightarrow -h$, and similarly, the part of the phase diagram where $\mu > U/2$ is symmetric to the shown part of the phase diagram with $\mu < U/2$ owing to the particle−hole symmetry. Solid lines demonstrate the results of numerical calculations. The length of each of the Hubbard chains (ladder legs) is $L = 40$ [8].

## 4. PROBLEMS OF THE EXPANSION AND COLLAPSE OF QUANTUM GASES

In [11–13], we studied the problem of the expansion of classical and quantum gases into vacuum using the symmetry approach. In the Gross–Pitaevskii approximation [14, 15], additional symmetries arise for quantum gases with a chemical potential $\mu \sim n^\nu$, where $\nu = 2/D$ and $D$ is the dimension of the space. For gas condensates of Bose atoms at temperatures $T \to 0$, this symmetry arises for two-dimensional systems. For $D = 3$ and, accordingly, $\nu = 2/3$, such a situation arises for a strongly interacting Fermi gas at low temperatures in the unitary limit [14]. The same symmetry for classical gases in the three-dimensional geometry takes place for monatomic gases with the heat capacity ratio $\gamma = 5/3$.

These properties of Bose and Fermi atoms were independently found in 1971 by Talanov [16] for the two-dimensional nonlinear Schrödinger equation (coinciding with the Gross–Pitaevskii equation [14, 15]), which describes the stationary self-focusing of light in media with Kerr nonlinearity, and by Anisimov and Lysikov [17] for classical gases. In the semiclassical limit, the Gross–Pitaevskii equation coincides with the equations describing the gas dynamics of the ideal gas with the heat capacity ratio $\gamma = 1 + 2/D$.

Self-similar solutions in this approximation describe the angular deformations of a gas cloud within Ermakov-type equations against the background of an expanding gas [18]. Such changes in the shape of the expanding cloud are observed in numerous experiments, both in the expansion of a gas after the intense laser irradiation of, e.g., a metal and in the expansion of quantum gases into vacuum.

The self-similar expansion of a strongly interacting Fermi gas from a cigar-shaped trap was observed in experiments reported in [19]. Figure 4, taken from [19], shows images of the expanding Fermi gas.

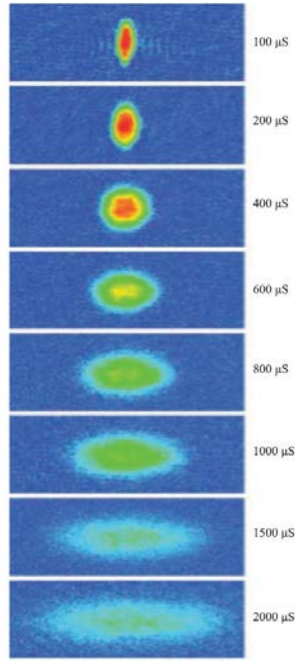

**Fig. 4.** Images of an expanding strongly interacting Fermi gas at the times indicated on the right. The initial cigar shape [19] is first transformed to a spherical shape and finally to a cylindrically symmetric pancake shape (from top to bottom).

Initially (at the time $t = 100$ μs), the gas cloud had a cigar shape; then, at $t = 600$ μs, it was almost spherical; and at the final stage, the cloud had a disk shape. The total observation time was equal to 2000 μs, which can be taken as a half period (or less) for the angular oscillations of the gas cloud shape, $t \leq t_{\text{osc}}/2$. All these stages qualitatively correspond to the self-similar solution. At the same time, the time dependence of the measured average radius squared of the gas cloud for different values of the inverse gas parameter $1/(a_s k_F)$, where $a_s$ is the $s$-wave scattering length and $k_F$ is the Fermi wave vector, is parabolic with a high accuracy, in full agreement with the relation for the mean square radius $R^2$ of the cloud, which follows from the virial theorem [20]:

$$2mN \frac{d^2 (R^2)}{dt^2} = 4E.$$

Here, $2m$ is the doubled (owing to the Cooper pairing) mass of a Fermi atom, $N$ is the total number of particles, and $E$ is the total energy. Integrating this relation, we obtain

$$2mNR^2 = 2m \int r^2 n d\mathbf{r} = 2Et^2 + C_1 t + C_2, \qquad (1)$$

where $n = |\Psi|^2$ is the density, $\Psi$ is the wavefunction of the condensate, and $C_1$ and $C_2$ are two additional integrals of motion. According to Eq. (1), the average size of the quantum gas cloud expanding into vacuum increases linearly with the time in the limit $t \to \infty$; i.e., the ballistic regime is reached.

As shown in [11–13], the Gross–Pitaevskii equation with $\mu \sim n^{2/D}$ allows exact three-dimensional self-similar solutions in the semiclassical approximation, which describe the anisotropic expansion of the gas into vacuum. In dimensionless variables, the density can be written in terms of self-similar variables $\xi_i = x_i/a_i$ ($i = 1, 2, 3$), where $a_i$ are the scaling parameters depending on the time $t$, as $n = f(\boldsymbol{\xi})/(a_1 a_2 a_3)$. Here, $\boldsymbol{\xi} = (\xi_1, \xi_2, \xi_3)$ and $f(\boldsymbol{\xi})$ is the spherically symmetric function, which depends only on $|\boldsymbol{\xi}|$ and is given by the formula

$$f(\boldsymbol{\xi}) = \left[ 1 - \frac{3\lambda}{10} \xi^2 \right]^{3/2},$$

where $\lambda$ is the positive constant determined from the initial conditions and vanishes at $|\boldsymbol{\xi}| > \sqrt{\frac{10}{3\lambda}}$.

The time dependence of the scaling parameters is determined from the solutions of the Newton equations

$$\ddot{a}_i = \frac{\partial U}{\partial a_i}, \ U(\mathbf{a}) = \frac{3\lambda}{2(a_1 a_2 a_3)^{2/3}} \qquad (2)$$

Here, $i = 1, 2, 3$ and $\mathbf{a} = (a_1, a_2, a_3)$. These equations belong to the so-called Ermakov-type systems [18] and are integrated because of the energy conservation law

$$E = \frac{1}{2} \dot{\mathbf{a}}^2 + U(\mathbf{a}). \qquad (3)$$

and Eq. (1), where $R^2 = |\mathbf{a}|^2 = 3a^2$, for $a_1 = a_2 = a_3 = a$, and the potential has the form

$$U(a) = 3\lambda \, / \, 2a^2. \tag{4}$$

In the spherically symmetric case, where the energy depends only on $|\mathbf{a}|$, it is quite trivial to find the solution. The size of the cloud $R$ increases linearly at $t \rightarrow \infty$; correspondingly, the expansion rate at large times tends to a constant value (ballistic regime), which is in dimensional variables equal to

$$v_\infty = \sqrt{E \, / \, mN}.$$

In the case of collapse [21–23], the expression for the cloud radius squared, which follows from the virial theorem, vanishes at a certain time $t = t_0$ (under a sufficient condition $H < 0$). This immediately implies that the size of the cloud near $t = t_0$ behaves as

$$R \propto (t_0 - t)^{1/2},$$

which corresponds to the regime of a weak self-similar collapse (see [21, 22]).

To analyze deformations of the cloud shape against the background of expanding gas (Fig. 4), it is necessary in (3) to pass to spherical coordinates, i.e., the radius $|\mathbf{a}|$ and the zenith and azimuth angles $\varphi$ and $\phi$. Owing to the virial theorem (1), the radial and rotational motions are separated, and the rotational dynamics is determined by the effective potential

$$U = \frac{3\lambda}{2^{1/3} \left( \cos^2 \phi \sin \phi \sin 2\varphi \right)^{2/3}} \cdot \frac{1}{3a^2}.$$

The minimum of this potential $\min U = \dfrac{3\lambda}{2a^2}$ achieved at

$$\sin \phi = 1 \, / \, \sqrt{3}, \, \varphi = \pi \, / \, 4, \tag{5}$$

corresponds to the spherically symmetric potential (4). The second derivatives of $U$ with respect to the angular variables at the point of minimum determine the main frequencies of oscillations. In the self-similar approximation, they are the same and their squares are given by the expression

$$\Omega_1{}^2 = \Omega_2{}^2 = 18\lambda.$$

Concluding of this section, we emphasize that the formation of large boson droplets (containing five or more particles) in magnetic and optical traps is typical of imbalanced Fermi–Bose mixtures of alkali isotopes Li-6 (fermions)–Li-7 (bosons) and K-40 (fermions)–Rb-87 (bosons) if the boson density exceeds the fermion one: $n_B > n_F$ (see [24]), as well as in Bose gases with resonance attraction between bosons. Large boson droplets in the case of a strong imbalance between the components $n_B \gg n_F$ were experimentally detected in the Fermi–Bose mixture K-40–Rb-87 with a large predominance of bosons in [25]. The threshold vanishing (with the number of bosons in the trap) of the local minimum in the boson energy and, as a result, the collapse in the boson subsystem in the imbalanced Fermi–Bose mixture with a predominance of bosons $N_B > N_F$ and with the attraction between fermions and bosons was analyzed in detail in theoretical works [26, 27].

In a balanced Fermi–Bose mixture with equal densities of the components, $n_B = n_F$, the formation of compound fermions $f_\sigma b$ and quartets $f_\uparrow b f_\downarrow b$ consisting of two fermions with opposite spin projections and two bosons is also possible as shown in [28, 29] and [30, 31], respectively. Large boson droplets were also experimentally observed in the Dy-164 Bose gas with a strong dipole–dipole attraction [32].

## 5. LARGE-WAVELENGTH CHARGE OSCILLATIONS IN MgB$_2$. ACOUSTIC PLASMONS. INHOMOGENEOUS STATES IN PLASMONIC SYSTEMS

The problem of the effective screened Coulomb interaction in a layered MgB$_2$ metal with an acoustic plasmon spectrum $\omega_{ac}^2 = B\sin^2(q_z d)$ in the $z$ direction perpendicular to the layers was formulated and qualitatively solved in [33] (see Fig. 5). We can clearly see in Fig. 5 that the spectrum is almost linear near the points $q_z d \to n\pi$, where $d$ is the interplanar spacing and $n = 0, 1, \ldots$.

In this case, the effective interaction in the momentum space has a singularity $V_{eff} \propto 1/|q_z - Q|$ at low frequencies in resonance with the acoustic plasmon frequency and finite wave vectors $q_z = Q = \pi/d$ equal to the "nesting" vector $Q$.

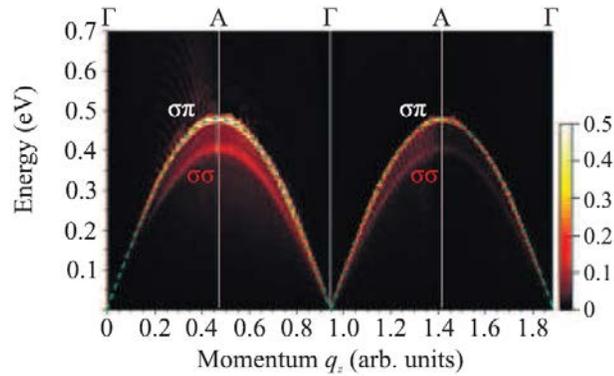

**Fig. 5.** Acoustic plasmon spectrum $\omega_{ac}^2 = B\sin^2(q_z d)$ in the layered MgB$_2$ metal along the $z$ direction perpendicular to the basal plane [33]. The spectrum near the $q_z d \to n\pi$ points, where $n = 0, 1, \ldots$ and $d$ is the interplanar spacing, is almost linear.

The careful transition to the real space gives a singular (oscillating) part of the effective interaction. This part decays as $\cos(Qz)/(Qz)^2$. Such a quadratic character of the decay is reminiscent of the Friedel oscillations corresponding to charge screening in a two-dimensional metal [34].

An interesting problem for the future is the possibility of Bose–Einstein condensation and superfluidity in weakly nonideal Bogoliubov gases of acoustic plasmons with the repulsive plasmon–plasmon interaction and with their collapse in the real space in the case of the attractive plasmon–plasmon interaction.

## 6. SUPERCONDUCTIVITY AND DROPLET FORMATION IN METALLIC HYDROGEN AND METAL HYDRIDES

We calculated the superconducting critical temperatures in homogeneous phases of metallic hydrogen and metal hydrides using our modified scheme of Eliashberg integral equations on the imaginary axis, taking into account corrections to the chemical potential [5, 35]. Numerical methods for solving the system of Eliashberg equations were also previously developed in [36–40]. In Fig. 6, we show the Eliashberg function of the phonon spectrum for one of the phases of metallic hydrogen under pressure.

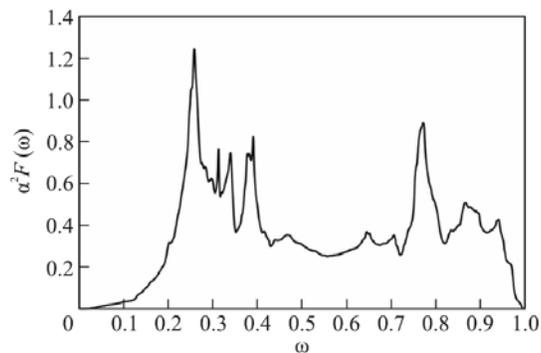

**Fig. 6.** Eliashberg phonon function for the $I4_1/amd$ phase of metallic hydrogen at a pressure of 500 GPa [38].

The temperature dependence of the order parameter is shown in Fig. 7. Although the difference between the results obtained for this phase of metallic hydrogen with the correction to the chemical potential of electrons [5, 35] and previously without it [38] is small, it can be quite significant for other substances. However, the inclusion of the correction to the chemical potential of electrons drastically (in our case, by two orders of magnitude) increases the calculation time.

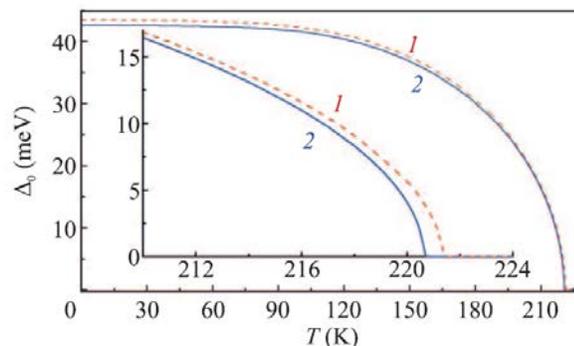

**Fig. 7.** Temperature dependence of the order parameter $\Delta(T)$ for the $I4_1/amd$ phase of metallic hydrogen at a pressure of 500 GPa calculated (dashed lines *1*) with and (solid lines *2*) without the correction to the chemical potential of electrons [35]. The inset presents the enlarged fragment of the plot near the superconducting transition temperature.

Similar results obtained at high pressures for other compounds of hydrogen were reported in [41]. For all substances, for which the Eliashberg functions are given in this work, we calculated $T_c$ by two methods (see Table 1): using the Allen–Dynes algorithm [42–44] (A–D-1 column) and the method of solving the Eliashberg equations from [5, 35] (El-1 column). The values of $T_c$ obtained in [41] using the modified Macmillan equation [42] in the Allen–Dynes approximation and by solving the Eliashberg equations are also shown in Table 1 (A–D-2 and El-2 columns).

**Table 1.** Superconducting transition temperatures of some hydrides at high pressures

| Compound | $P$ (GPa) | $T_c$ (K) | | | |
|---|---|---|---|---|---|
| | | A–D-1 | A–D-2 | El-1 | El-2 |
| $LaBeH_8$ | 50 | 194 | 167 | 192 | 191 |
| $LaBH_8$ | 70 | 160 | 144 | 159 | 160 |
| $LaAlH_8$ | 100 | 145 | 130 | 145 | 144 |
| $CaBeH_8$ | 210 | 307 | 254 | 303 | 302 |
| $CaBH_8$ | 100 | 241 | 212 | 239 | 238 |
| $YBeH_8$ | 100 | 250 | 215 | 248 | 249 |
| $SrBH_8$ | 150 | 206 | 163 | 204 | 200 |

We can see in Table 1 that, first, the results of our calculations and the results from [41] nearly coincide with an accuracy of 1 K (the discrepancy is obviously due to the error in digitizing the plots of the Eliashberg functions) and provide a very good approximation for the superconducting transition temperature (no worse than 2 K for most materials). On the contrary, the use of the modified Macmillan equation in the Allen–Dynes approximation underestimates $T_c$ by several dozen degrees.

### 6.1. Formation of Metallic Droplets near the Boundary of the First-Order Phase Transition between Metallic and Molecular Hydrogen

As a natural generalization of the results presented in this work, we plan to extend our model for the Eliashberg equations to take into account the effect of impurities, the multiband nature of the compounds under study [3, 5, 35], and the possibility of the formation of inhomogeneous states (formation of droplets) that arise at approaching the first-order phase transition between liquid and crystalline metallic hydrogen.

We are also going to take into account the possibility of the formation of two Bose–Einstein condensates in the system, the first of which is a Bose–Einstein condensate of Cooper pairs in the electron subsystem, and the second one is a diproton Bose–Einstein condensate in the ionic component (much like the situation in neutron stars). Qualitative reasons for the coexistence of two Bose–Einstein condensates of Cooper pairs and diprotons in low-dimensional phases of metallic hydrogen (such as the planar and filamentary phases [24, 45]) were revealed in our works [46, 47].

Let us now consider the possibility of droplet formation in metallic hydrogen. In Fig. 8, we show the $T-P$ phase diagram of molecular and atomic metallic hydrogen at high pressures. The phase diagram contains four phases of molecular and metallic hydrogen that are important for us, namely, two phases of solid (crystalline) molecular and metallic hydrogen at high pressures and low temperatures and two phases of liquid molecular and metallic hydrogen at high pressures and high temperatures [48].

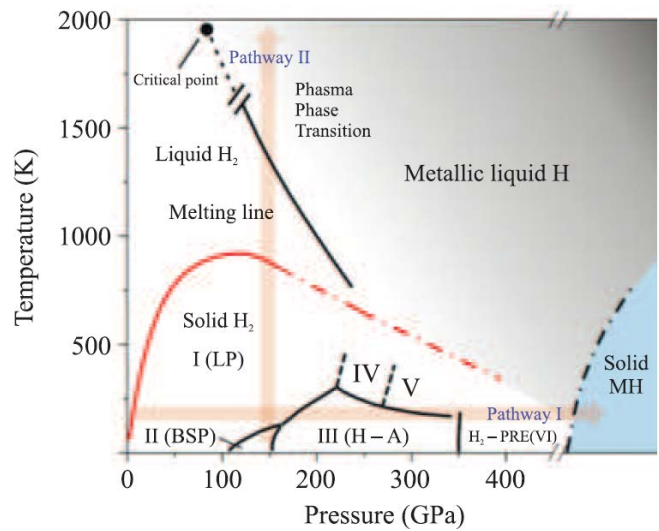

**Fig. 8.** $T-P$ phase diagram of molecular and atomic metallic hydrogen at high pressures. The phase diagram near the phase boundary between molecular and metallic hydrogen contains four phases that are essential for us, namely, two phases of solid (crystalline) molecular and metallic hydrogen at high pressures and low temperatures and two phases of liquid molecular and metallic hydrogen at high pressures and high temperatures [48].

In this case, the phase transition from the crystalline to the liquid phase for both atomic metallic hydrogen and molecular (insulating) hydrogen is a first-order phase transition, which is characterized by the formation of inhomogeneous droplet structures. The formation of liquid droplets and small "ice particles" around positive and negative ions is well known in the physics of liquid and solid helium [49]. Similarly, one can expect the formation of metallic hydrogen droplets in an insulating host of molecular hydrogen. The experimental confirmation of this idea is provided by the shock-wave experiments [50] with deuterium systems.

## 7. COLLAPSE OF THE FANO RESONANCE IN THE AHARONOV–BOHM RING [51] WITH A TOPOLOGICAL SUPERCONDUCTOR BRIDGE AS A TEST FOR THE NONLOCALITY OF MAJORANA MODES

The concepts of spatially separated Fermi and Bose systems and of nontrivial superconductivity induced in them can also be implemented in a restricted microcontact geometry with two or four quantum dots [52–54].

Figure 9 shows the rhombic geometry with four quantum dots, where two Fermi arms (in fact, two effective Fermi bands), which are described by tunneling Hamiltonians for Fermi quasiparticles, are connected by a superconducting bridge containing paired bosonic states [55–57]. In this case, a topologically nontrivial superconductivity can be induced in the superconducting bridge [55].

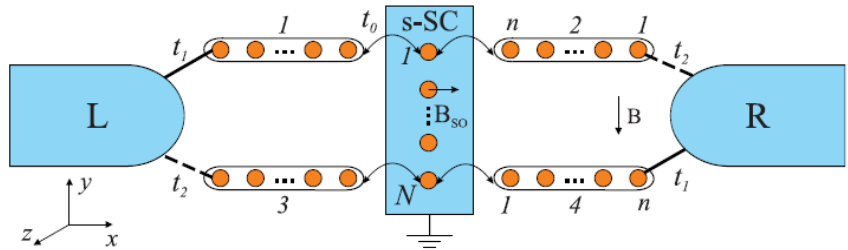

**Fig. 9.** Aharonov–Bohm ring [51] with the superconducting bridge in a topologically nontrivial state with the asymmetry in the tunnel parameters between the ring arms and contacts. The Aharonov–Bohm ring is used as an interference device for observing the Fano resonances [58].

In this context, S.V. Aksenov and M.Yu. Kagan [57] solved in 2020 the problem of the collapse of a narrow Fano resonance [58] in the conductance of the system caused by the nonlocality of the Majorana state [59]. We emphasize that one of the main features of Majorana states, which stimulates great interest in these excitations in solid-state systems, is their nonlocal nature.

In [57], we considered an interference device in the form of an asymmetric Aharonov–Bohm ring with the arms connected by a bridge perpendicular to them in the form of a one-dimensional topological superconductor. The device is asymmetric because the tunneling parameters $t_1$ and $t_2$ between the contacts and the two arms of the ring are different.

It is shown that, in contrast to the symmetric geometry considered in our work [56], new Fano resonances arise in the conductance of the asymmetric device. It is found that their width is directly proportional to the energy of the nonlocal state of the superconductor with a minimum energy equal to the binding energy of a pair of edge Majorana excitations.

In other words, the more the probability density is concentrated at the edges, the narrower is the Fano resonance. As a result, in the limiting case of two noninteracting Majorana fermions (very long topological superconducting bridge), this feature in the conductance disappears (see Fig. 10).

This statement can be illustrated [57] within the spinless ring model of the Kitaev type [60], where the Majorana chain with an even number of sites is used as a bridge. It is shown that the observed effect is related to an increase in the multiplicity of degeneracy in the structure with zero energy at the singular point of the Kitaev model and to a decrease in the overlap of the wavefunctions of Majorana fermions with the growing number of sites (in fact, the length) in the chain, which leads to the collapse of the narrow Fano resonance.

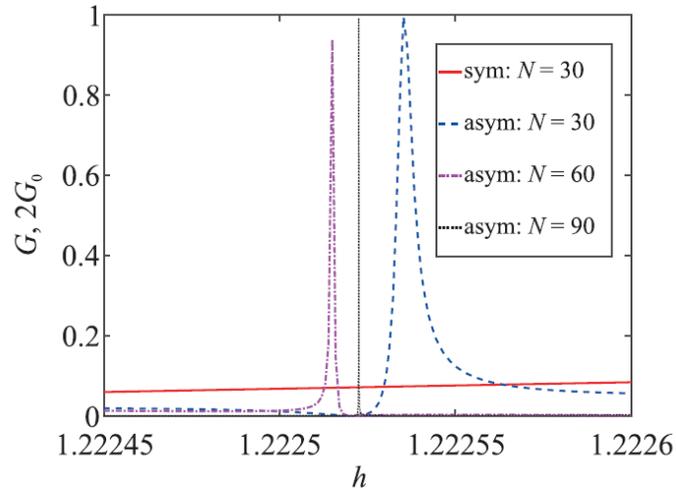

**Fig. 10.** Collapse of the narrow Fano resonance with an increase in the degree of nonlocality of the Majorana modes in the superconducting bridge.

The additional study of a particular case of only half of an interferometer (for example, when the superconducting bridge in Fig. 9 is connected only to the left wires) demonstrated that the overlap of the Majorana wavefunctions that form the zero-energy Bogoliubov excitation leads to a doubling of the period of Aharonov−Bohm oscillations [61].

## 8. CONCLUSIONS

To summarize, a number of superconducting and superfluid systems, in which droplets or clusters of the order parameter can emerge in the hosts involving normal unpaired states, have been described in the framework of a spatially separated Fermi–Bose mixture. These systems include superconducting bismuth oxides BaKBiO, imbalanced Fermi gases and Fermi–Bose mixtures in magnetic traps, and two-dimensional systems with a low electron density described by the two-dimensional Hubbard model with a strong local on-site attraction and strong diagonal disorder. The formation of droplets is also typical of collapsing exciton [62] and plasmon systems with the attractive exciton–exciton and plasmon–plasmon interactions, as well as for metallic hydrogen in shock-wave experiments near the onset of the first-order phase transition between the liquid and crystalline phases.

The results reported here may be useful for applications in superconducting nanoelectronics and for the implementation of superconducting qubits in granular media.

FUNDING

This work was supported by the Russian Foundation for Basic Research (project no. 20-02-00015). M.Yu. Kagan acknowledges the support of the National Research University Higher School of Economics (Program of Basic Research).

REFERENCES

1. A.P. Menushenkov, K.V. Klementev, A.V. Kuznetsov, and M.Yu. Kagan, J. Exp. Theor. Phys. **93**, 615 (2001).
2. A.P. Menushenkov, A.V. Kuznetsov, K.V. Klementiev, and M.Yu. Kagan, J. Supercond. Nov. Magn. **29**, 701 (2016).
3. M.Yu. Kagan, K.I. Kugel, and A.L. Rakhmanov, Phys. Rep. **916**, 1 (2021).
4. M.Yu. Kagan and E.A. Mazur, J. Exp. Theor. Phys. **132**, 596 (2021).
5. E.A. Mazur, R.Sh. Ikhsanov, and M.Yu. Kagan, J. Phys.: Conf. Ser. **2036**, 012019 (2021).
6. Y. Shin, M.W. Zwierlein, C.H. Schunck, A. Schirotzek, and W. Ketterle, Phys. Rev. Lett. **97**, 030401 (2006).
7. W. Ong, C. Cheng, I. Arakelyan, and J.E. Thomas, Phys. Rev. Lett. **114**, 110403 (2015).
8. E.A. Burovski, R.Sh. Ikhsanov, A.A. Kuznetsov, and M.Yu. Kagan, J. Phys.: Conf. Ser. **1163**, 012046 (2019).
9. P. Fulde and R.A. Ferrell, Phys. Rev. A **135**, 550 (1964).
10. A.I. Larkin and Yu.N. Ovchinnikov, Sov. Phys. JETP **20**, 762 (1964).
11. E.A. Kuznetsov, M.Yu. Kagan, and A.V. Turlapov, Phys. Rev. A **101**, 041612 (2020).
12. E.A. Kuznetsov and M.Yu. Kagan, Theor. Math. Phys. **202**, 399 (2020).
13. E.A. Kuznetsov and M.Yu. Kagan, J. Exp. Theor. Phys. **132**, 704 (2021).
14. L.P. Pitaevskii, Phys. Usp. **51**, 603 (2008).
15. E.P. Gross, Nuovo Cim. **20**, 454 (1961).
16. V.I. Talanov, JETP Lett. **11**, 199 (1971).
17. S.I. Anisimov and Yu.I. Lysikov, Prikl. Mat. Mekh. **34**, 926 (1970).
18. V.P. Ermakov, in *Lectures on Integration of Differential Equations* (Univ. Tipogr., Kiev, 1880) [in Russian].
19. K.M. O'Hara, S.L. Hemmer, M.E. Gehm, S.R. Granade, and J. E. Thomas, Science **298**, 2179 (2002).
20. S.N. Vlasov, V.A. Petrishchev, and V.I. Talanov, Izv. Vyssh. Uchebn. Zaved., Radiofiz. **14**, 1353 (1971) [in Russian].
21. V.E. Zakharov and E.A. Kuznetsov, Sov. Phys. JETP **64**, 773 (1986).
22. V.E. Zakharov and E.A. Kuznetsov, Phys. Usp. **55**, 535 (2012).
23. V.E. Zakharov, Sov. Phys. JETP **35**, 908 (1972).
24. E.G. Brovman, Yu. Kagan, A. Kholas, and V.V. Pushkarev, JETP Lett. **18**, 160 (1973).
25. G. Modugno, G. Roati, F. Riboli, F. Ferlaino, R.J. Brecha, and M. Inguscio, Science **297**, 2240 (2002).
26. S.T. Chui and V.N. Ryzhov, Phys. Rev. A **69**, 043607 (2004).
27. S.T. Chui, V.N. Ryzhov, and E.E. Tareyeva, JETP Lett. **80**, 274 (2004).

28. M.Yu. Kagan, I.V. Brodsky, D.V. Efremov, and A.V. Klaptsov, Phys. Rev. A **70**, 023407 (2004).

29. M.Yu. Kagan, A.V. Klaptsov, I.V. Brodsky, R. Combescot, and X. Leyronas, Phys. Usp. **49**, 1079 (2006).

30. A.V. Turlapov and M.Yu. Kagan, J. Phys.: Condens. Matter **29**, 383004 (2019).

31. M.Yu. Kagan and A.V. Turlapov, Phys. Usp. **62**, 215 (2019).

32. I.F. Barbur, H. Kadan, M. Schmitt, M. Wenzel, and T. Pfau, Phys. Rev. Lett. **116**, 215301 (2016).

33. V.M. Silkin, A. Balassis, P.M. Eschenique, and E.V. Chulkov, Phys. Rev. B **80**, 054521 (2009).

34. M.Yu. Kagan, V.A. Mitskan, and M.M. Korovushkin, Phys. Usp. **58**, 733 (2015).

35. R.Sh. Ikhsanov, E.A. Mazur, and M.Yu. Kagan, Izv. Ufim. Nauch. Tsentra RAN **1**, 49 (2023) [in Russian].

36. R. Szczesniak, Acta Phys. Polon. A **109**, 179 (2006).

37. A.P. Durajski, Sci. Rep. **6**, 38570 (2016).

38. N.A. Kudryashov, A.A. Kutukov, and E.A. Mazur, JETP Lett. **104**, 460 (2016).

39. I.A. Kruglov, D.V. Semenok, H. Song, R. Szcześniak, I.A. Wrona, R. Akashi, E.M.M. Davari, D. Duan, C. Tian, A.G. Kvashnin, and A.R. Oganov, Phys. Rev. B **101**, 024508 (2020).

40. O.V. Dolgov, R.K. Kremer, J. Kortus, A.A. Golubov, and S.V. Shulga, Phys. Rev. B **72**, 024504 (2005).

41. Z. Zhang, T. Cui, M.J. Hutcheon, A.M. Shipley, H. Song, M. Du, V.Z. Kresin, D. Duan, C.J. Pickard, and Y. Yao, Phys. Rev. Lett. **128**, 047001 (2022).

42. P.B. Allen and R.C.A. Dynes, Phys. Rev. B **12**, 905 (1975).

43. F. Marsiglio and J.P. Carbotte, in *Superconductivity,* vol. 1. *Conventional and Unconventional Superconductors* (Springer, Berlin, 2008), p. 73.

44. J.P. Carbotte, Rev. Mod. Phys **62**, 1027 (1990).

45. E.G. Brovman, Yu. Kagan, and A. Kholas, Sov. Phys. JETP **34**, 1300 (1972).

46. M.Yu. Kagan, JETP Lett. **103**, 728 (2016).

47. M.Yu. Kagan and A. Bianconi, Condens. Matter **4**, 51 (2019).

48. M. Houtput, J. Tempere, and I.F. Silvera, Phys. Rev. B **100**, 134106 (2019).

49. I.M. Khalatnikov, *An Introduction to the Theory of Superfluidity* (Nauka, Moscow, 1965; CRC, Boca Raton, FL, 2000).

50. M.D. Knudson, M.P. Desjarlais, A. Becker, R.W. Lemke, K.R. Cochrane, M.E. Savage, D.E. Bliss, T.R. Mattsson, and R. Redmer, Science **348**, 1455 (2015).

51. Y. Aharonov and D. Bohm, Phys. Rev. **115**, 485 (1959).

52. M.Yu. Kagan, V.V. Val'kov, and S.V. Aksenov, Phys. Rev. B **95**, 035411 (2017).

53. M.Yu. Kagan, V.V. Val'kov, and S.V. Aksenov, J. Magn. Magn. Mater. **440**, 15 (2017).

54. M.Yu. Kagan and S.V. Aksenov, JETP Lett. **107**, 493 (2018).

55. V.V. Val'kov, M.S. Shustin, S.V. Aksenov, A.O. Zlotnikov, A.D. Fedoseev, V.A. Mitskan, and M.Yu. Kagan, Phys. Usp. **65**, 2 (2022).

56. S.V. Aksenov, M.Yu. Kagan, and V.V. Val'kov, J. Phys.: Condens. Matter **31**, 225301 (2019).

57. S.V. Aksenov and M.Yu. Kagan, JETP Lett. **111**, 286 (2020).

58. U. Fano, Phys. Rev. **124**, 1866 (1961).

59. E. Majorana, Nuovo Cim. **5**, 171 (1937).

60. A.Yu. Kitaev, Phys. Usp. **44** (Suppl.), 131 (2001).

61. S.V. Aksenov, J. Phys.: Condens. Matter **34**, 255301 (2022).

62. L.V. Keldysh, Phys. Usp. **60**, 1180 (2017).